# Gate-tunable large second-order and third-order nonlinear Hall effect in topological insulator $BiSbTeSe_2$

Fengyi Guo[1], Shuai Zhang[1,2,3,*], Xiubing Li[1], Zheng Dai[1], Congcong Li[1], Yufan Ju[1], Ziqi Wang[1], Cheng Bin[1,4], Zhe Ying[3], Fengqi Song[1,2,3,*]

[1] National Laboratory of Solid State Microstructures, Collaborative Innovation Center of Advanced Microstructures, Jiangsu Physical Science Research Center, and School of Physics, Nanjing University, Nanjing 210093, China

[2] Institute of Atom Manufacturing, Nanjing University, Nanjing 211800, China

[3] Jiangsu Provincial Key Laboratory of Atomic Level Manufacturing, Nanjing Institute of Atomic Scale Manufacturing, Nanjing 211800, China

[4] School of Physics and Electronic Information, Jiangsu Second Normal University, Nanjing 211200, China

*Corresponding authors. S.Z. (szhang@nju.edu.cn), F.S. (songfengqi@nju.edu.cn)

**Abstract**

The nonlinear Hall effect (NLHE) holds significant promise for next-generation electronic device applications, and it is essential to explore and engineer a strong and highly tunable NLHE. In this work, we report large second-order and third-order NLHEs in dual-gated $BiSbTeSe_2$ devices. Both the second-order and third-order NLHE are highly gate-tunable. The second-order NLHE exhibits a higher generation efficiency than those reported in previous topological insulator systems, and it can be stably observed up to room temperature. Meanwhile, the generation efficiency of the third-order NLHE is comparable to previously reported values. The scaling analysis indicates that the second-order NLHE is consistent with contributions from skew scattering and a term compatible with either the Berry curvature dipole or side jump, whereas the third-order NLHE is consistent with contributions from the Berry connection polarizability tensor and higher-order skew scattering. Our work establishes a promising experimental platform for the development of NLHE-based devices.

## I. INTRODUCTION

The Hall effect and its extended family hold a significant position in condensed matter physics. It refers to the phenomenon where a transverse voltage is generated perpendicular to the direction of an applied longitudinal current[1, 2]. The conventional Hall effect requires time-reversal symmetry (TRS) breaking. In contrast, the second-order nonlinear Hall effect (NLHE) occurs under the condition of inversion symmetry breaking without requirement for TRS breaking[3-5]. Under an a.c. current, the *n*-th-harmonic component of the Hall voltage $V_{xy}^{n\omega}$ is conventionally used to characterize the *n*-th-order nonlinear Hall response. The second-order NLHE can be attributed to the Berry curvature dipole (BCD)[3, 6-15], the quantum metric dipole[16-19] and the side jump or skew scattering[5, 20-32]. Recent studies have demonstrated that mechanisms such as the Berry connection polarizability tensor (BPT)[33-39], Berry curvature quadrupole[40-44], as well as high-order side jump or skew scattering[45-47] can induce the third-order NLHE. The third-order NLHE originating from the BPT is independent of both inversion symmetry and time-reversal symmetry[48]. Both the second-order and third-order NLHE were first observed in $WTe_2$[6, 33]. Recently, the two types of NLHE have also been realized in $MoTe_2$[33, 49], $TaIrTe_4$[14, 35], $MnBi_2Te_4$[17, 18], $(SnS)_{1.17}(NbS_2)_3$[39], $Fe_5GeTe_2$[46], $ZrTe_5$[32, 37], $Cd_3As_2$[38], bismuth[31] and graphene[8, 28, 47]. Currently, the study of the NLHE is concentrated on obtaining stronger nonlinear signals[7], achieving higher operating temperatures[14, 19, 20, 35], and realizing gate voltage modulation[18, 20, 33, 38], which are critical for advancing the practical device applications of the NLHE, including high-frequency rectifiers, frequency multipliers, and nonlinear logic devices.

Topological insulators (TIs) have attracted considerable attention owing to the topological surface states (TSSs) with spin-momentum locking[50, 51]. Since the bulk of the TIs preserves spatial inversion symmetry, the inversion-symmetry-broken TSS transport channel provides an important electronic platform for the second-order NLHE.[22, 25, 27, 32, 37]. Meanwhile, theoretical studies have indicated that TSSs can also host the third-order NLHE[48]. Recently, scattering-induced second-order NLHE has been reported in $Bi_2Se_3$[22] and $Bi_2Te_3$[25, 27]. In addition, the second-order NLHE has also been observed in the magnetic TI $MnBi_{2x}Te_{3x+1}$ family[17, 18, 26]. However, the third-order NLHE has not been observed in strong three-dimensional TIs. It has only been reported in weak TI $ZrTe_5$[37], as well as in magnetic TIs under an external magnetic field[41]. Although the NLHE can persist at relatively high temperatures, the tunability of the NLHE by gate voltage is rather weak[22, 25, 27], which limits its practical applications. $BiSbTeSe_2$ (BSTS) is a representative strong three-dimensional TI, in which bulk carrier contribution is significantly suppressed, transport properties are dominated by TSS, and it features tunable Dirac carriers[52] . Thus, BSTS provides an excellent platform for studying the NLHE.

In this work, we observe both the second-order and third-order NLHE in BSTS thin-flake devices. Both the second-order and third-order NLHE are highly gate-tunable. Meanwhile, the scaling analysis indicates that the second-order NLHE is consistent with contributions from skew scattering and a term compatible with either the BCD or side jump, whereas the third-order NLHE is consistent with contributions from the BPT and higher-order skew scattering. Experimental results show that the generation efficiency of the second-order NLHE in this system is significantly higher than that reported in previously studied (magnetic) TIs, and it remains stable up to 300 K. The signal intensity of the third-order NLHE is comparable to that in existing works. Moreover, effective modulation of the NLHE in BSTS can be achieved by tuning the Fermi level ($E_F$) via gate voltage. It benefits the design of nonlinear electronic devices.

## II. EXPERIMENTAL METHODS

The high-quality $BiSbTeSe_2$ crystals were grown by melting high-purity elements of Bi, Sb, Te, Se with a molar ratio of 1:1:1:2. The $BiSbTeSe_2$ flakes were mechanically exfoliated onto a highly *p*-doped Si substrate with 300-nm-thick $SiO_2$. Gold electrodes were fabricated by electron beam lithography and electron beam evaporation. The thickness of the $BiSbTeSe_2$ flakes was measured by atomic force microscopy (Cypher S). The *h*-BN and graphite were transferred on top of the $BiSbTeSe_2$ by dry transfer.

An a.c. excitation current $I^{\omega}$ with a frequency of 31 Hz was applied using a current source (Keithley 6221). The longitudinal and Hall voltages at the first-, second- and third-harmonic frequencies were detected using lock-in amplifiers (Stanford Research Systems SR830). These voltage components are denoted as $V_{xx}^{\omega}$, $V_{xx}^{2\omega}$, and $V_{xx}^{3\omega}$ for the longitudinal channel and $V_{xy}^{\omega}$, $V_{xy}^{2\omega}$, and $V_{xy}^{3\omega}$ for the Hall channel, respectively. In particular, $V_{xy}^{2\omega}$ and $V_{xy}^{3\omega}$ represent the second- and third-harmonic responses of the Hall voltage and are used to characterize the second-order and third-order NLHEs, respectively. The harmonic-voltage amplitudes are root-mean-square (RMS) values directly recorded by the lock-in amplifiers rather than amplitudes extracted from harmonic fitting. The $I^{\omega}$ value is the RMS value of the a.c. current. A Keithley 2400 source meter was used to apply the gate voltage. All transport measurements were performed in a Cryomagnetics cryostat.

## III. RESULTS and DISCUSSIONS

The BSTS devices were fabricated via electron-beam lithography and electron beam evaporation. The hexagonal boron nitride (*h*-BN) and graphite were transferred onto the BSTS by dry-transfer technique. The inset in Fig. 1(a) is the structure of the dual-gated BSTS device and transport measurement configuration. Graphite/h-BN serves as the top gate ($V_{tg}$), and the *p*-doped-Si/$SiO_2$ layer serves as the back gate ($V_{bg}$). The thickness of the BSTS measured by atomic force microscopy is about 10.7 nm (see Fig. S1 in Supplementary Materials[53]). Figure 1(a) illustrates the current ($I^{\omega}$) dependent voltage at 10 K with a frequency of 31 Hz. The linear $I^{\omega}$-$V_{xx}$ curve indicates good ohmic contacts. Meanwhile, the Hall voltage $V_{xy}$ is vanishingly small. The temperature ($T$) dependent longitudinal resistance $R_{xx}$ is shown in Fig. 1(b). As the temperature decreases, the resistance first increases and then decreases, indicating the dominant role of TSSs at low temperatures[54]. The current dependent second-order nonlinear Hall voltage ($V_{xy}^{2\omega}$) with reversing the source–drain contacts and Hall contacts is shown in Fig. 1(c). The inset illustrates the measurement configuration. It is found that the sign of $V_{xy}^{2\omega}$ reverses with the current direction, while the magnitude remains nearly unchanged. This behavior is distinct from the conventional Hall effect and excludes the contribution of heating effects[10]. For the third-order nonlinear Hall voltage ($V_{xy}^{3\omega}$), as shown in Fig. 1(d), the sign of $V_{xy}^{3\omega}$ remains unchanged. The two $I^{\omega}$-$V_{xy}^{3\omega}$ curves almost coincide with each other. In Figs. 1(e) and 1(f), the NLHE response as a function of current overlaps well at different frequencies, ruling out the spurious capacitive coupling effect[24]. The contact-junction effect and heating effect are also ruled out (see Figs. S13–S16 in Supplementary Materials[53]).

The gate-dependent NLHE is shown in Fig. 2. Figures 2(a)-2(c) exhibit the dual-gate mapping of $R_{xx}$, $V_{xy}^{2\omega}$, and $V_{xy}^{3\omega}$ at $T$=10 K and $I^{\omega}$ = 5 μA, respectively. Figure 2(d) shows the dependence of the resistance on the top-gate voltage $V_{tg}$ when $V_{bg}$ = 13 V. The resistance reaches a maximum at $V_{tg}$ = 6.4 V, corresponding to the charge neutrality point (CNP) of the BSTS device. Within the same gate voltage range, the evolutions of $V_{xy}^{2\omega}$, and $V_{xy}^{3\omega}$ are presented in Figs. 2(e) and 2(f), respectively. In Fig. 2(e), the second-order nonlinear Hall voltage evolves with gate voltage from the peak, through zero with a sign reversal, to the valley. In contrast, under the same gate modulation conditions, the third-order nonlinear Hall voltage exhibits distinct behaviors (Fig. 2(f)). Specifically, in addition to reaching a peak[22, 38], sign reversals of the $V_{xy}^{3\omega}$ signal occur at $V_{tg}$ = 5.8 V and $V_{tg}$ = 7.2 V, on either side of the CNP. While in the region far away from the CNP, the second-order and third-order nonlinear Hall voltage approach zero.

The pronounced response of $V_{xy}^{2\omega}$ and $V_{xy}^{3\omega}$ near the CNP indicates that the NLHE in BSTS is closely related to the band structure and depends strongly on the $E_F$. When $E_F$ approaches the CNP, the bulk is insulating, while the TSSs within the band gap dominate the transport. When gate voltages are far away from the CNP, $E_F$ shifts into the bulk valence or conduction band, and the bulk states mainly contribute to transport[48]. The pronounced second-order and third-order nonlinear Hall responses near the CNP, together with their suppression far from the CNP, suggest that the TSSs play an important role in both nonlinear responses.

To elucidate the distinct behaviors of the third-order NLHE, we present the gate evolution of the NLHE measured far from the resistance maximum ( see Fig. S3 in Supplementary Materials[53]). With back gate modulation, the third-order nonlinear Hall voltage exhibits a peak as a function of the top gate. In contrast, top gate modulation leads to a valley of the third-order nonlinear Hall voltage. Meanwhile, the second-order nonlinear Hall voltage undergoes a sign reversal at the CNP. It is known that both the top and bottom surface contribute to the third-order NLHE, with their contributions of opposite signs, $V_{xy}^{3\omega} = V_{xy_\mathrm{t}}^{3\omega} + V_{xy_\mathrm{b}}^{3\omega}$. The third-order nonlinear Hall voltage of a single surface follows[55, 56] $V_{xy_\mathrm{t(b)}}^{3\omega} \propto \frac{1}{|E_{F_\mathrm{t(b)}}|}$, and the total response is the sum of each surface, $V_{xy}^{3\omega} \propto \frac{|E_{F_\mathrm{b}}|-|E_{F_\mathrm{t}}|}{|E_{F_\mathrm{b}}||E_{F_\mathrm{t}}|}$. The core features of the third-order NLHE in Fig. 2 and Fig. S3 can then be explained. When $E_{F_\mathrm{b}}$ is far from the CNP, $E_{F_\mathrm{t}}$ can gradually shift from far below to far above the CNP with the top gate tuning. As shown in Fig. S3(c), when $V_{\mathrm{bg}}$ = 1 V and $V_{\mathrm{tg}}$ = 8.5 V, $\left|E_{F_\mathrm{t}}\right| \approx 0$ is satisfied. At this time, the dominant contribution to the third-order NLHE comes from the top surface, $V_{xy}^{3\omega} \propto \frac{1}{|E_{F_\mathrm{t}}|}$ following, which results in the peak in Fig. S3(c). By the same token, the emergence of the valley in Fig. S3(f) can also be explained. Near the resistance maximum, $\left|E_{F_\mathrm{t}}\right| \approx \left|E_{F_\mathrm{p}}\right| \approx 0$ is satisfied. At this point, the third-order NLHE is co-dominated by the top and bottom surfaces, $V_{xy}^{3\omega} \propto \frac{|E_{F_\mathrm{b}}|-|E_{F_\mathrm{t}}|}{|E_{F_\mathrm{t}}|^2}$, where$\left|E_{F_\mathrm{b}}\right| - \left|E_{F_\mathrm{t}}\right|$ is

a first-order infinitesimal and $|E_{F_t}|^2$ is a second-order infinitesimal. This accounts for the peak at the resistance maximum. On both sides of the resistance maximum, $|E_{F_t}| \approx |E_{F_b}| \neq 0$, the third-order NLHE is also co-dominated by the top and bottom surfaces. However, $V_{xy}^{3\omega} \propto \frac{|E_{F_b}|-|E_{F_t}|}{|E_{F_t}|^2} \approx 0$, which further explains the emergence of the sign reversal behavior on both sides of the resistance maximum. Meanwhile, although the second-order NLHE contributions from the top and bottom surfaces partially cancel each other[22], the second-order NLHE on each surface individually reverses sign at the CNP[20]. As a result, the overall second-order nonlinear Hall voltage still exhibits sign reversal at the CNP.

Then, we select a typical set of gate voltage in device D1, specifically ($V_{tg}$, $V_{bg}$) = (6 V, 8.5 V) near the CNP to study the temperature dependent $V_{xy}^{2\omega}$ and $V_{xy}^{3\omega}$. The current dependent second-order and third-order nonlinear Hall voltages at different temperatures are shown in Figs. 3(a) and 3(d), respectively. Near the CNP, the NLHE is significantly enhanced. The second-order nonlinear Hall voltage on both sides of the CNP is about ten times larger than that under the gate-free condition (see Fig. S4 in Supplementary Materials[53]), while the third-order nonlinear Hall signal becomes detectable up to 50 μV from nearly negligible. The second-order nonlinear Hall voltage persists up to 300 K, whereas the third-order nonlinear Hall voltage gradually decreases with increasing temperature and vanishes at a lower temperature (see Figure S9 in Supplementary Materials[53] for the temperature dependent NLHE of device D3).

Notably, in contrast to the BCD-induced NLHE, which keeps increasing continuously after sign reversal induced by $E_F$ variation[14], the BSTS device in this work exhibits a distinctly different temperature dependent behavior. The second-order nonlinear Hall response undergoes a sign change at ~50 K, peaks in magnitude near 100 K, and then progressively decreases to vanishingly small values, yet remains detectable up to 300 K. These behaviors are also verified in another device (D4) (see Fig. S10 in Supplementary Materials[53]). Within the temperature evolution of the above characteristic points, the third-order NLHE decays rapidly with increasing temperature. This behavior originates from the gradual weakening of the dominant contribution of TSSs to charge transport as temperature increases. Meanwhile, bulk transport gradually becomes dominant, which ultimately reduces the third-order nonlinear Hall voltage to 0.15 μV at 250 K.

Since $V_{xy}^{n\omega}$ is proportional to the *n*-th power of current, we define $\eta^{(n)}=V_{xy}^{n\omega}/V_{xx}^{n}$ to characterize the NLHE intensity, referred to as the NLHE generation efficiency. Figures 3(b) and 3(e) display the temperature dependence of $\eta^{(n)}$ at fixed gate voltages of ($V_{\mathrm{tg}}$, $V_{\mathrm{bg}}$) = (6 V, 8.5 V). The generation efficiencies were calculated directly using $V_{xy}^{2\omega}$, $V_{xy}^{3\omega}$, and $V_{xx}^{\omega}$ measured at the same current of $I^{\omega}$ = 5 μA. At low temperatures, $|\eta^{(2)}|$ reaches up to 0.3 $\mathrm{V}^{-1}$, and $|\eta^{(3)}|$ can reach 5 $\mathrm{V}^{-2}$.

Then, we discuss the origin of the NLHE. Similar to the AHE[2], the scaling law can be used to distinguish different origins of the NLHE[10, 33]. We find that $\eta^{(n)}$ exhibits a linear dependence on the square of conductance $(G_{xx})^2$, $\eta^{(n)}=A(G_{xx})^2+B$, where *A* reflects the skew-scattering contribution, while *B* represents an intrinsic BCD contribution or side jump[10, 33] (see Table S1 in the Supplementary Materials [53] for the fitting results). The temperature dependence is shown in Figs. 3(c) and 3(f). Figure 3(c) shows the scaling analysis for the second-order NLHE, where the gray, red, and blue curves correspond to the temperature ranges of 10–80 K, 100–150 K, and 175–300 K, respectively. It can be seen that the gray curve shows the largest slope, and the variation of the fitted slopes with temperature suggests an evolution of the skew scattering related contribution (see in Fig. S5(a) in Supplementary Materials[53]). The scaling analysis indicates that the second-order NLHE is consistent with contributions from skew scattering and a term compatible with either the BCD or side jump. Moreover, $|\eta^{(2)}|$ in BSTS is three orders of magnitude larger than that in $Bi_2Se_3$. The NLHE compared to recent works in other TI systems[15, 17, 18, 22, 25, 27, 32] is shown in Fig. 4 (see Table S2 in Supplementary Materials[53]), which demonstrates the high performance of BSTS.

The third-order NLHE in nonmagnetic systems can receive contributions from both the BPT and higher-order skew scattering following the linear scaling relation[33, 37], $\eta^{(3)}=A(G_{xx})^2+B$. A linear scaling behavior is shown in Fig. 3(f) (see details in Fig. S5(b) in Supplementary Materials[53]). The third-order NLHE is consistent with contributions from the BPT and higher-order skew scattering. However, the third-order NLHE has not been observed in strong three-dimensional TIs, because crystal symmetry plays a crucial role in observing the third-order NLHE. The third-order NLHE is forbidden and vanishes when the two-dimensional system possesses $C_{3v}$, $C_{6v}$, $D_3$, $D_{3h}$, $D_{3d}$, or $D_6$ symmetries[34, 48]. In the strong three-dimensional TIs represented by $Bi_2Se_3$[22] and $Bi_2Te_3$[25, 27], the third-order NLHE is symmetry-forbidden and unobservable. The observation of third-order NLHE in TIs requires the breaking of $C_3$ symmetry[48].

To further rationalize the origin of the third-order NLHE, we fabricated a 12-electrode disk-shaped device (D5) (see Figs. S11 and S12 in the Supplementary Materials[53]). At 15 K, the angular dependences of the first-order longitudinal voltage and the second-order and third-order nonlinear Hall voltages on $\theta$ are shown in Figs. 5(a)–5(c). The first-order longitudinal voltage and the third-order nonlinear Hall voltage exhibit distinct twofold angular dependences, whereas the second-order nonlinear Hall voltage shows a onefold angular dependence. For the third-order NLHE, the twofold angular dependence is compatible with the symmetry requirements for a BPT contribution. The angular dependences can be phenomenologically described by the nonlinear susceptibility tensors allowed under a possible reduction of the effective surface symmetry from $C_{3v}$ to $C_s$. Nevertheless, the present measurements establish only a reproducible twofold angular response and do not uniquely identify the origin of the effective symmetry reduction. Doping- or chemical-substitution-induced modifications of the electronic and transport properties may contribute[57], although chemical substitution alone would not necessarily produce a macroscopic twofold axis unless the associated compositional or structural inhomogeneity possesses a preferred in-plane orientation. Preferentially oriented local structural features, such as defects, terraces, flake edges, or thickness variations, may also contribute, but their roles are not directly established in the present devices. Fabrication-induced strain or residual stress was also considered because strain can substantially modify the Dirac surface states of Bi-based topological insulators[58]. However, no intentional strain was applied, and the present measurements provide no direct evidence for appreciable uniaxial strain. Strain or residual stress is therefore unlikely to be the dominant origin. Moreover, the reproducibility of the twofold angular response in another 12-electrode disk-shaped BSTS device (D2) indicates that a particular contact configuration, device geometry, or nonuniform current spreading is unlikely to account for the observed behavior (see Figs. S7 and S8 in the Supplementary Materials [53]). Determining the exact microscopic origin requires further local structural characterization, which is beyond the scope of this work.

## IV. CONCLUSIONS

In summary, we observe both the second-order and third-order NLHE in dual-gated BSTS thin-flake devices. The NLHE signals can be modulated by gate voltage, and the second-order NLHE remains stable at room temperature. Meanwhile, the scaling analysis indicates that the second-order NLHE is consistent with contributions from skew scattering and a term compatible with either the BCD or side jump, whereas the third-order NLHE is consistent with contributions from the BPT and higher-order skew scattering. The superior properties of the BSTS, including high generation efficiency, gate tunability and room-temperature feasibility, indicate that it possesses excellent prospects for the application of high-performance nonlinear electronic devices.

## ACKNOWLEDGMENTS

We acknowledge funding from the National Natural Science Foundation of China (Grant Nos. 12374043, 92580203, 12025404, 12504053, T2221003), the National Key R&D Program of China (No. 2022YFA1402404), the Natural Science Foundation of Jiangsu Province (Nos. BK20240166, BK20243013, BK20233001), the Fundamental and Interdisciplinary Disciplines Breakthrough Plan of the Ministry of Education of China (No. JYB2025XDXM411), and the Fundamental Research Funds for the Central Universities (Nos. KG202501, KG202603).

## DATA AVAILABILITY

The data that support the findings of this article are not publicly available upon publication because it is not technically feasible and/or the cost of preparing, depositing, and hosting the data would be prohibitive within the terms of this research project. The data are available from the authors upon reasonable request.

## Figures and captions

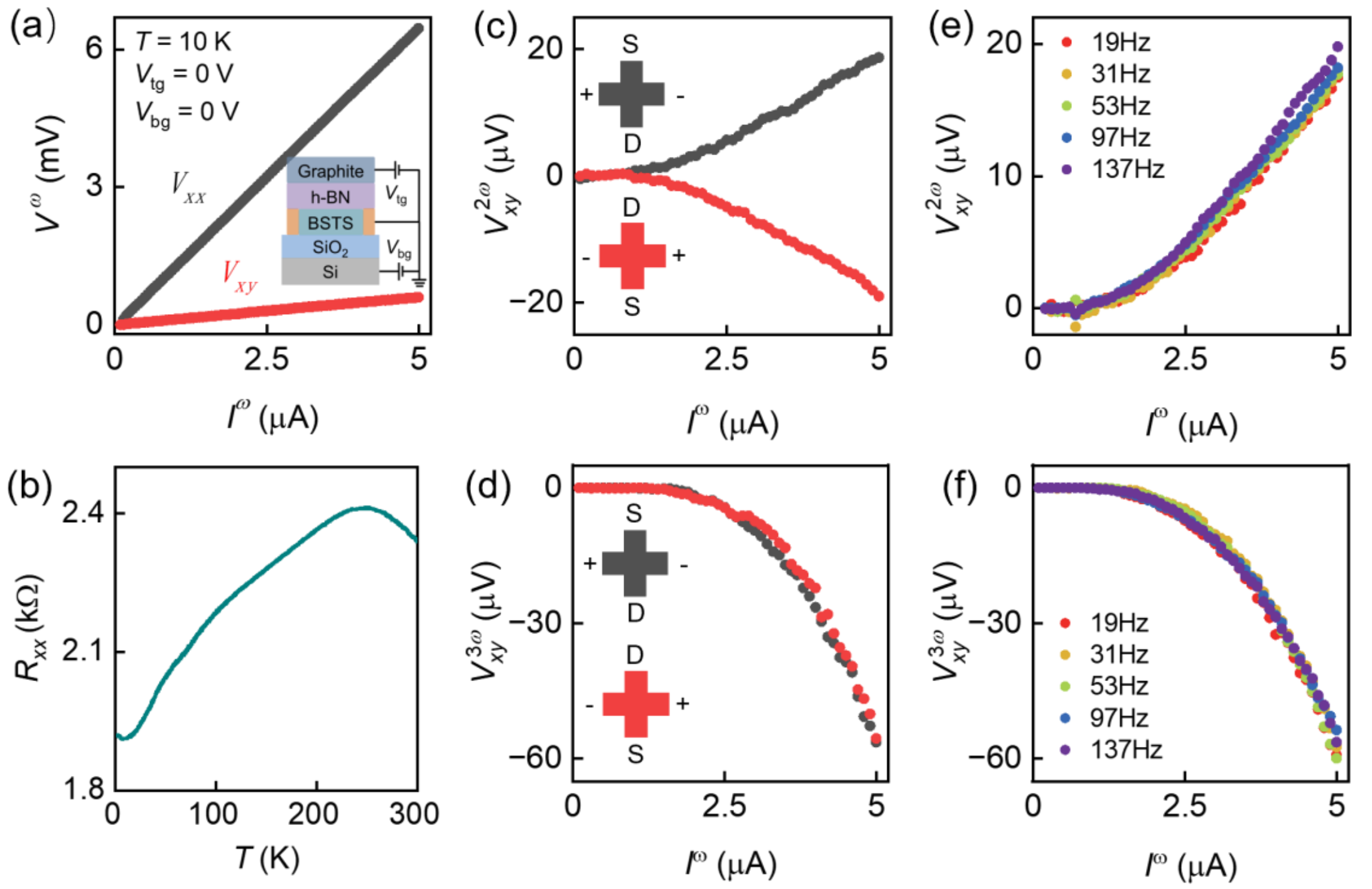

FIG. 1. BSTS Hall device and the nonlinear Hall properties (D1). (a) Current dependent first-order longitudinal voltage $V_{xx}$ and Hall voltage $V_{xy}$ at 10 K. The inset is the structure of the dual-gated BSTS device. $Si/SiO_2$ and graphite/*h*-BN work as back and top gates, respectively. The thickness of BSTS is 10.7 nm. (b) Temperature dependent resistance $R_{xx}$. (c) The $I^{\omega}$-$V_{xy}^{2\omega}$ curves for the forward current (black curve) and backward current (red curve), while the voltage leads connection is also reversed when the current direction changes from forward to backward. (d) The $I^{\omega}$-$V_{xy}^{3\omega}$ curves for the forward current (black curve) and backward current (red curve). (e)-(f) Current dependent second-order and third-order nonlinear Hall voltages at different frequencies.

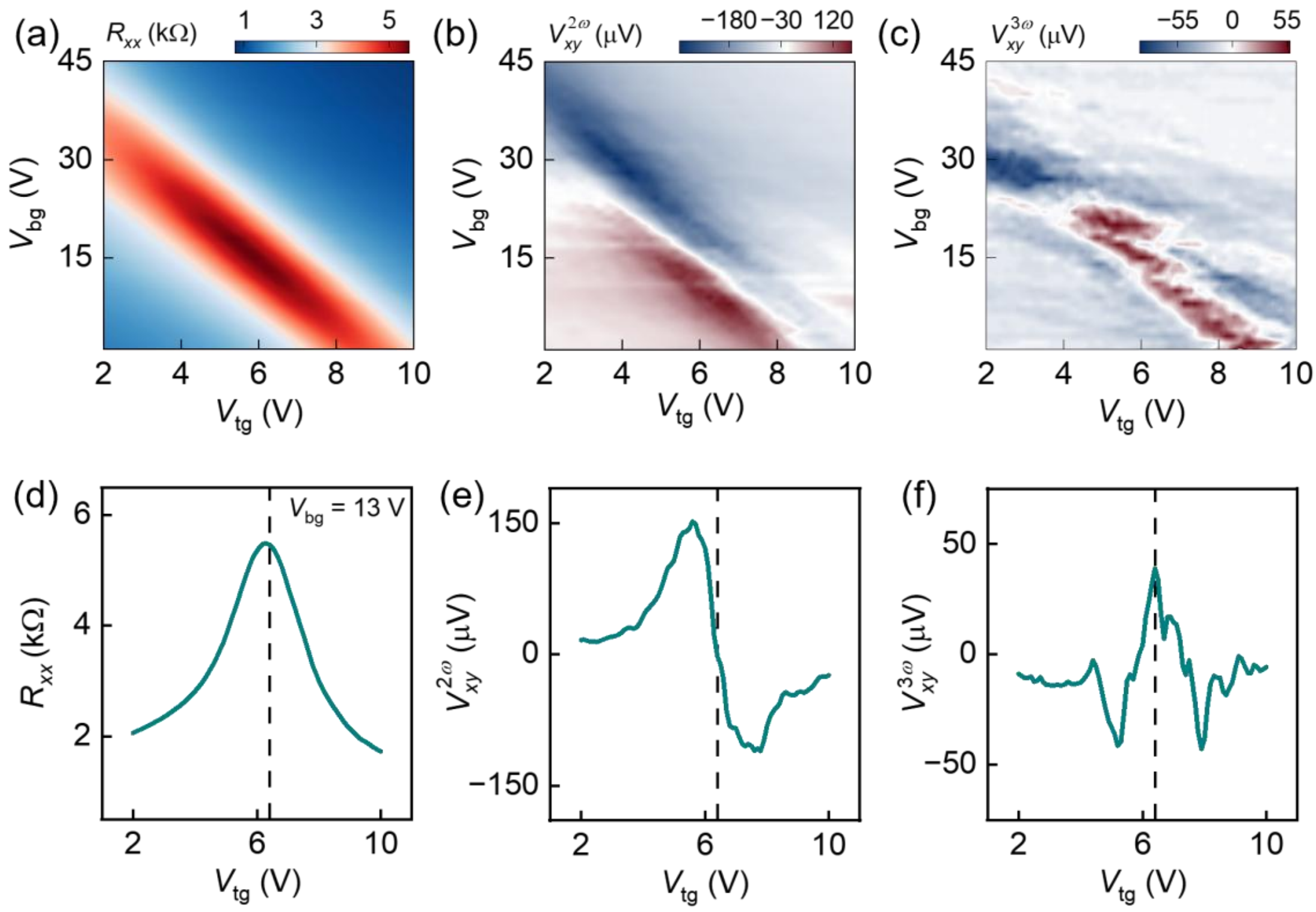

FIG. 2. Gate-dependent nonlinear Hall effect of BSTS device (D1). (a) The mapping of longitudinal resistance $R_{xx}$ as a function of top gate voltage ($V_{tg}$) and bottom gate voltage ($V_{bg}$) at $T = 10$ K. (b)-(c) The mapping of second-order and third-order nonlinear Hall voltages ($V_{xy}^{2\omega}$ and $V_{xy}^{3\omega}$). (d) The top gate voltage $V_{tg}$ dependence of the resistance $R_{xx}$ at a fixed back gate voltage $V_{bg} = 13$ V. (e)-(f) The top gate voltage $V_{tg}$ dependence of the second-order and third-order nonlinear Hall voltages ($V_{xy}^{2\omega}$ and $V_{xy}^{3\omega}$) under the same back gate voltage.

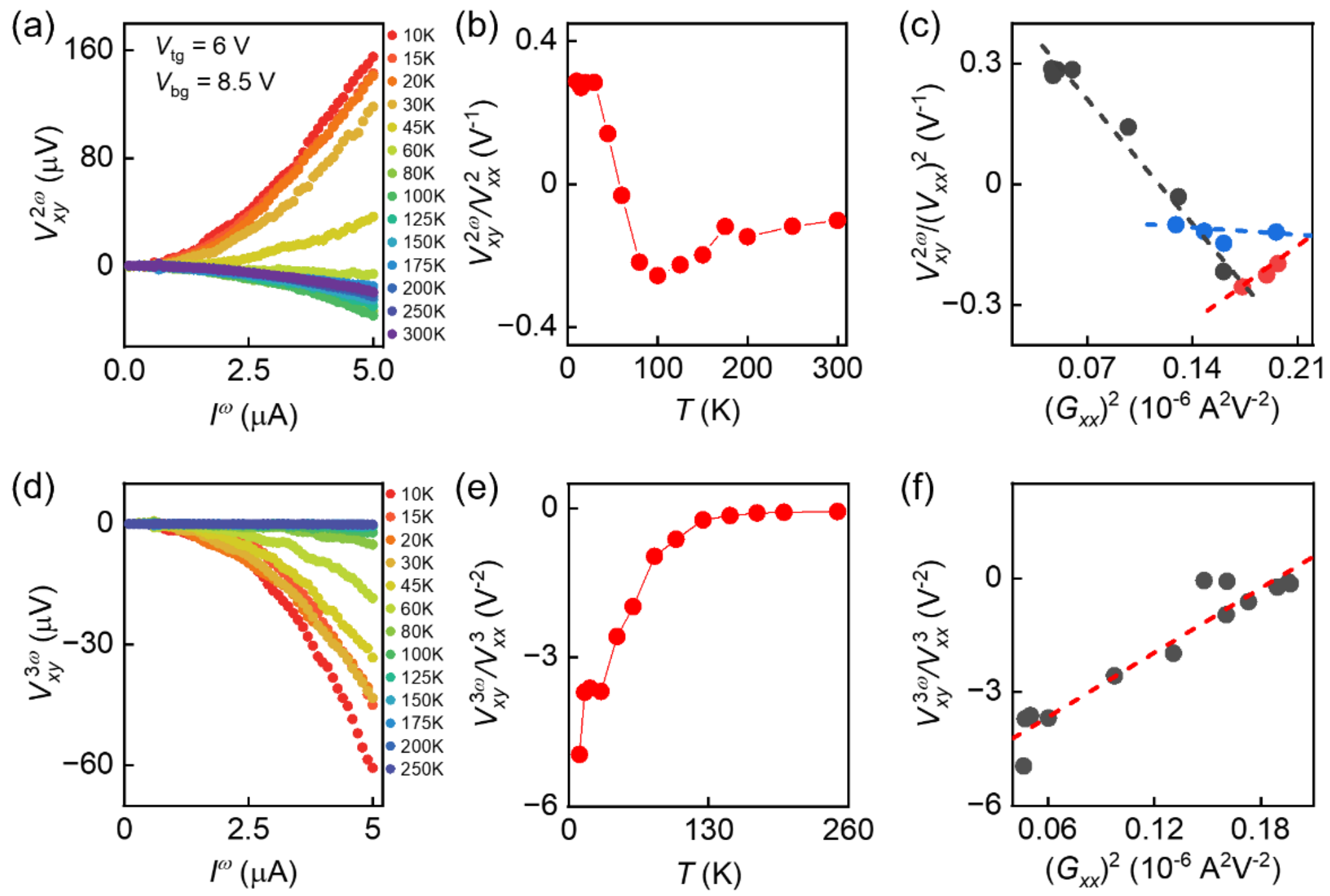

FIG. 3. Second-order and third-order nonlinear Hall voltage of BSTS Hall device at ($V_{tg}$, $V_{bg}$) = (6 V, 8.5 V) (D1). (a) The $I^{\omega}$-$V_{xy}^{2\omega}$ curves at temperatures from 10 K to 300 K. (b) Temperature dependence of the second-order nonlinear Hall generation efficiency. The generation efficiency $\eta^{(2)}$ was calculated using $V_{xy}^{2\omega}$ and $V_{xx}^{\omega}$ at $I^{\omega}$ = 5 μA. (c) $\eta^{(2)}$ plotted as a function of the square of the longitudinal conductance ${G_{xx}}^2$. The dashed lines are linear fittings. (d) The $I^{\omega}$-$V_{xy}^{3\omega}$ curves at temperatures from 10 K to 250 K. (e) Temperature dependence of the third-order nonlinear Hall generation efficiency. The generation efficiency $\eta^{(3)}$ was calculated using $V_{xy}^{3\omega}$ and $V_{xx}^{\omega}$ at $I^{\omega}$= 5 μA. (f) $\eta^{(3)}$ plotted as a function of the square of ${G_{xx}}^2$. The dashed line is a linear fitting.

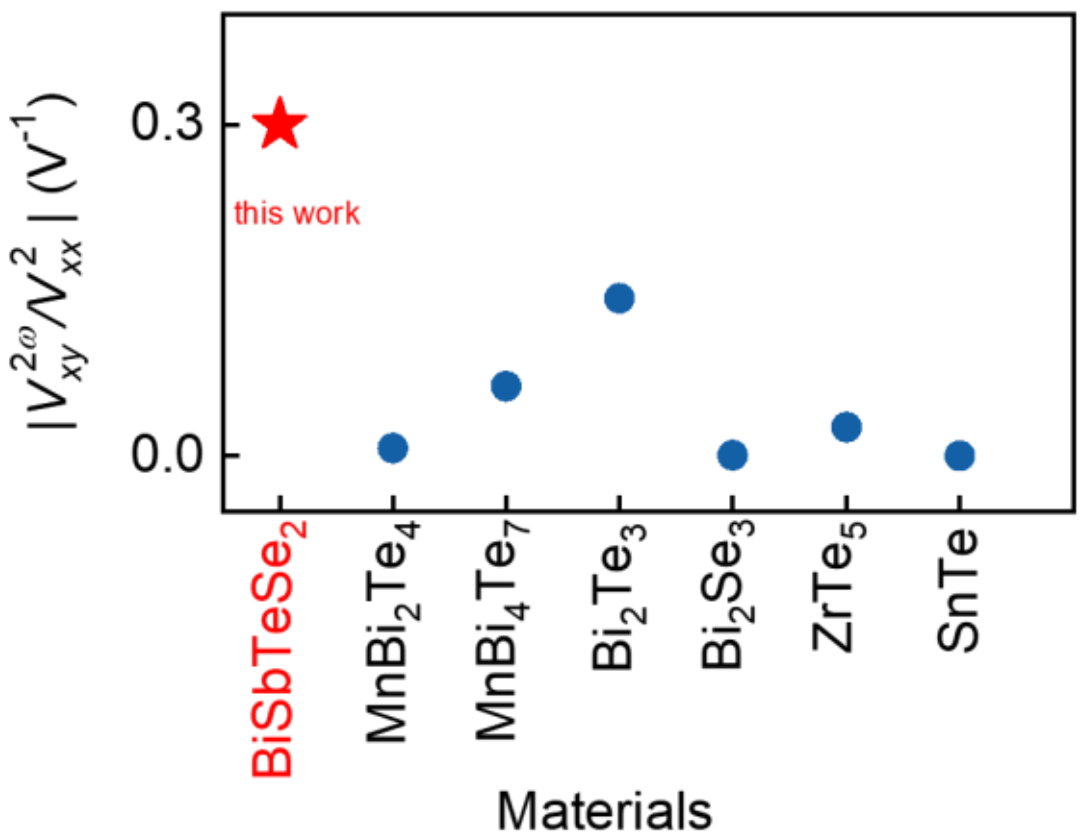

FIG. 4. The second-order nonlinear Hall generation efficiency compared to recent studies on topological insulator systems.

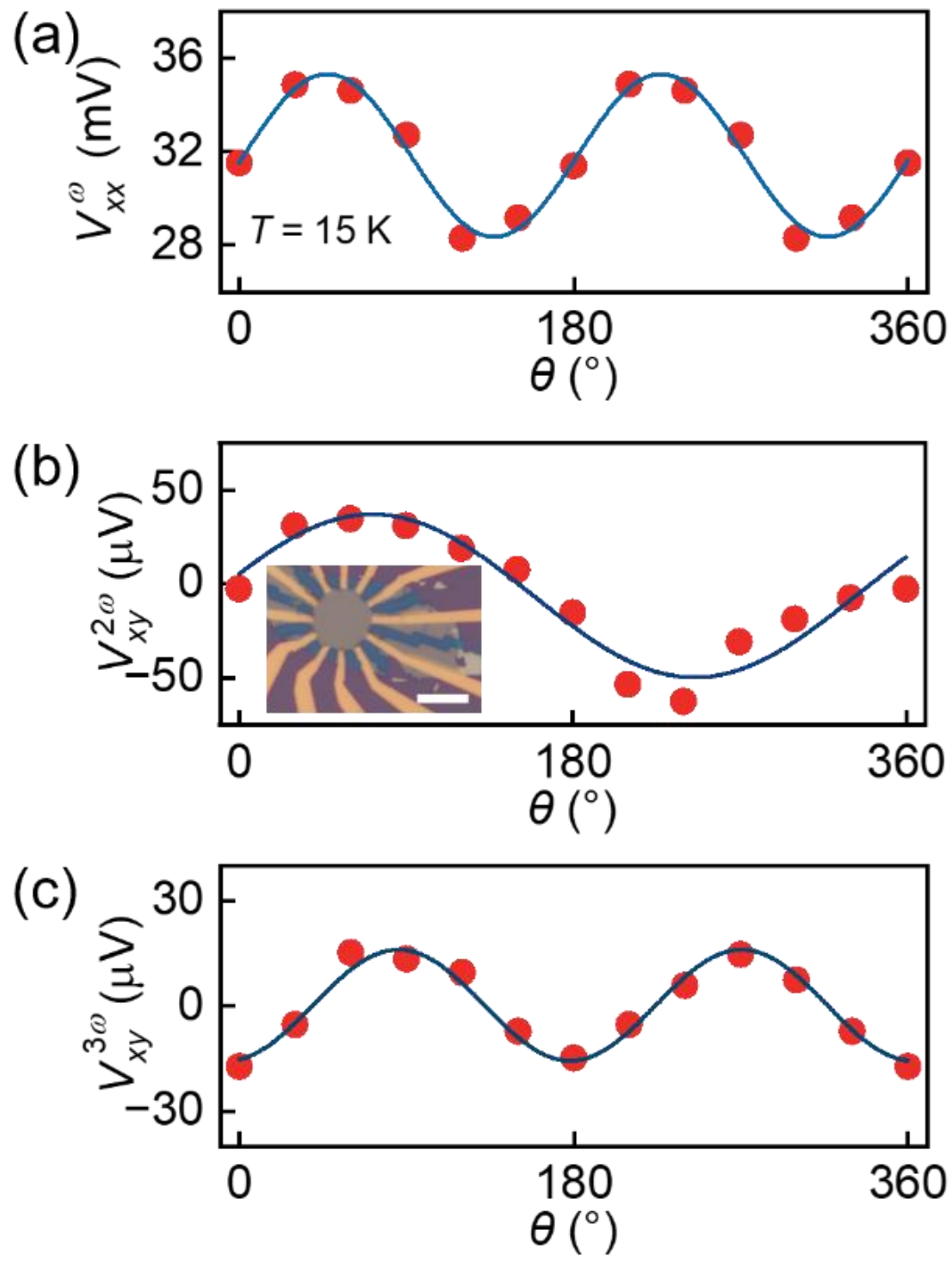

FIG. 5. Angle-dependent nonlinear Hall effect in disk-shaped BSTS device (D5). (a)-(c) Angular dependence of the first-order longitudinal voltage $V_{xx}^{\omega}$, second-order nonlinear Hall voltages $V_{xy}^{2\omega}$ and third-order nonlinear Hall voltages $V_{xy}^{3\omega}$ at $T$ = 15 K and $I^{\omega}$ = 5 μA. Inset in (b) is the optical image of the disk-shaped BSTS device (D5).